\documentclass{article}

\usepackage[psamsfonts]{amssymb}
\usepackage{amsmath}
\usepackage{cite}
\usepackage{eufrak}
\usepackage{bm}
\usepackage{graphicx}
\def\e{\EuFrak{e}}

\def\d{\mathrm{d}}
\def\dim{\mathrm{dim}}
\def\id{\mathbf{1}}
\def\reg{\mathrm{reg}}
\def\c{\star}
\def\ir{\mathrm{i}}
\def\vo{\mathrm{vol}(G)}
\begin{document}
\begin{titlepage}
\noindent{\large\textbf{Field theory amplitudes in a space with SU(2) fuzziness}}

\vskip 1.5 cm

\begin{center}
{Haniyeh Komaie-Moghaddam{\footnote {haniyeh.moghadam@gmail.com}}\\
Amir~H.~Fatollahi{\footnote {ahfatol@gmail.com}} \\
Mohammad~Khorrami{\footnote {mamwad@mailaps.org}} } \vskip 10 mm
\textit{ Department of Physics, Alzahra University, Tehran
1993891167, Iran. }

\end{center}

\vspace{0.5cm}

\begin{abstract}
\noindent The structure of transition amplitudes in field theory
in a three-dimensional space whose spatial coordinates are
noncommutative and satisfy the SU(2) Lie algebra commutation relations
is examined. In particular, the basic notions for constructing the
observables of the theory as well as subtleties related to the
proper treatment of $\delta$ distributions (corresponding to
conservation laws) are introduced. Explicit examples are given for
scalar field theory amplitudes in the lowest order of
perturbation.
\end{abstract}
\end{titlepage}
\section{Introduction}
Recently much attention has been paid to the formulation and study
of field theories on noncommutative spaces. The motivation is
partly the natural appearance of noncommutative spaces in some
areas of physics, a recent one occurring in string theory. In
particular, it has become clear that the longitudinal directions
of D-branes in the presence of a constant B-field background
appear to be noncommutative, as seen by the ends of open strings
\cite{9908142,99-2,99-3,99-4}. In this case the coordinates
satisfy the canonical relation
\begin{equation}\label{fk.1}
[\widehat x_a,\widehat x_b]=\ir\,\theta_{a\, b}\,\id,
\end{equation}
in which $\theta$ is an antisymmetric constant tensor and $\id$
represents the unit operator. Although due to the presence of the
background field, it might seem as if a Poincar\'{e} invariant
interpretation of field theories on canonical noncommutative
spaces is not possible, it has been shown that a twisted version
of Poincar\'{e} symmetry can be introduced as the alternative
symmetry of field theories on canonical spaces \cite{tureanu}. The
theoretical and phenomenological implications of possible
noncommutative coordinates have extensively been studied
\cite{reviewnc}.

One direction to extend studies on noncommutative spaces is to
consider spaces for which the commutators of the coordinates are
not constants. Examples of this kind are the cases with a
$q$-deformed plane and noncommutative cylinder ($S^1\times
\mathbb{R}$) \cite{chai}. It is shown that, while the
ultraviolet (UV) behavior of the theory in a $q$-deformed case is
worse than an ordinary plane, the theory on a noncommutative
cylinder, contrary to its commutative version, appears to be
UV-finite \cite{chai}. Another example of this kind is the so
called $\kappa$-Poincar\'{e} algebra, in which the
noncommutativity is introduced between spatial directions and
time, that is \cite{majid,ruegg}
\begin{align}\label{fk.3}
[\widehat x_a,\widehat t\;]=&\frac{\ir}{\kappa}\,\widehat x_a,\cr
[\widehat x_a,\widehat x_b]=&~0,
\end{align}
where $\kappa$ is a constant. The formulation of quantum field theories on
this kind of spaces has been studied in \cite{amelino,kappa}.

In the noncommutative cylinder and $\kappa$-Poincar\'{e}
cases mentioned above the noncommutativity is involved by
the time direction. Other interesting examples are the models in
which the (dimensionless) \textit{spatial} positions operators satisfy the
commutation relations of a Lie algebra \cite{wess,grosse1}:
\begin{equation}\label{fk.2}
[\widehat x_a,\widehat x_b]=f^c{}_{a\, b}\,\widehat x_c,
\end{equation}
where the $f^c{}_{a\,b}$'s are structure constants of a Lie
algebra. One example of this kind is the algebra SO(3), or SU(2).
A special case of this is the so called fuzzy sphere
\cite{madore,presnaj}, where an irreducible representation of the
position operators is used that makes the Casimir operator of the
algebra, $(\widehat x_1)^2+(\widehat x_2)^2+(\widehat x_3)^2$, a
multiple of the identity operator (a constant, hence the name
sphere). One can consider the square root of this Casimir operator
as the radius of the fuzzy sphere. This is, however, a
noncommutative version of a two-dimensional space (sphere).
Different aspects of field theories on the fuzzy sphere, including
the fate of the UV-divergences of the Euclidean theory, the
structure of UV/IR mixing, as well as topologically nontrivial
field configurations have already been examined
\cite{grosse2,presnaj2,grosse3}.

In a previous work \cite{0612013} a model was introduced in which
the representation was not restricted to an irreducible one;
instead the whole group was employed. In particular, the regular
representation of the group, which contains all representations,
was considered. As a consequence in such models one is dealing
with the whole space, rather than a 2-dimensional sub-space as in
the case of fuzzy sphere. The space of the corresponding momenta
is an ordinary (commutative) space and is compact if and only if
the group is compact. In fact, one can consider the momenta as the
coordinates of the group. So a by-product of such a model would be
the elimination of any UV-divergence in any field theory
constructed on such a space. One important implication of the
elimination of the ultraviolet divergences, as we shall see in
more detail later, would be that there will not remain  place for
the so called UV/IR mixing effect \cite{MRS}, which is known as a
common phenomenon one expects to  be going to face in a  models
with canonical noncommutativity, the algebra (\ref{fk.1}). In
\cite{0612013} the basic ingredients for calculus on a linear fuzzy
space, together with basic notions for a field theory on such a
space, including Lagrangian and elements for a perturbation
theory, were introduced. The models based on the regular
representations of SU(2) and SO(3) were treated in more detail,
giving the explicit form of the tools and notions introduced in
their general form.

In the present work the aim is to examine the structure of
amplitudes coming from a field theory based on a space with SU(2)
fuzziness. In particular, we introduce the basic elements by which
one can compute the matrix elements corresponding to the
transition between initial and final states. The contribution
entailed in a perturbative expansion of the amplitudes are
presented in the lowest order (tree level) for a self-interacting
scalar field theory.

The scheme of this paper is the following. In section 2, a brief
review is given of the calculus and field theory on noncommutative
spaces of the  Lie algebra type, so that the present paper is
self-contained. In section 3 the basic elements of transition
matrix elements are introduced and discussed. Explicit examples
are presented to show how thiese things work. Section 4 is devoted
to our conclusion.

\section{Basic notions}
\subsection{Calculational tools}
For a compact group $G$, there is a unique measure $\d U$ (up to a
multiplicative constant) with the invariance properties
\begin{align}\label{fk.4}
\d (V\,U)&=\d U,\nonumber\\
\d (U\,V)&=\d U,\nonumber\\
\d (U^{-1})&=\d U,
\end{align}
for an arbitrary element $V$ of the group. These mean that this
measure is invariant under the left-translation,
right-translation, and inversion. This measure, the
(left-right-invariant) Haar measure, is unique up to a
normalization constant, which defines the volume of the group:
\begin{equation}\label{fk.5}
\int_G\d U=\vo.
\end{equation}
Using this measure, one constructs a vector space as follows.
Corresponding to each group element $U$ an element $\e(U)$ is
introduced, and the elements of the vector space are linear
combinations of these elements:
\begin{equation}\label{fk.6}
f:=\int\d U\;f(U)\,\e(U),
\end{equation}
The group algebra is this vector space, equipped with the
multiplication
\begin{equation}\label{fk.7}
f\,g:=\int\d U\,\d V\; f(U)\,g(V)\,\e(U\,V),
\end{equation}
where $(U\,V)$ is the usual product of the group elements. $f(U)$
and $g(U)$ belong to a field (here the field of complex numbers).
It can be seen that if one takes the central extension of the
group U(1)$\times\cdots\times$U(1), the so-called Heisenberg
group, with the algebra (\ref{fk.1}), the above definition results
in the well-known star product of two functions, provided $f$ and
$g$ are interpreted as the Fourier transforms of the functions.

So there is a correspondence between functionals defined on the
group, and the group algebra. The definition (\ref{fk.7}) can be
rewritten as
\begin{align}\label{fk.8}
(f\,g)(W)=&\int\d V\;f(W\,V^{-1})\,g(V),\nonumber\\
=&\int\d U\;f(U)\,g(U^{-1}\,W).
\end{align}
Using  Schur's lemmas, one proves the so called grand
orthogonality theorem, which states that there is an orthogonality
relation between the matrix functions of the group:
\begin{equation}\label{fk.9}
\int\d U\; U_\lambda{}^a{}_b\,U^{-1}_\mu{}^c{}_d=
\frac{\vo}{\dim_\lambda}\,\delta_{\lambda\,\mu}\,\delta^a_d\,\delta^c_b,
\end{equation}
where $U_\lambda$ is the matrix of the element $U$ of the group in
the irreducible representation $\lambda$, and $\dim_\lambda$ is
the dimension of the representation $\lambda$. Exploiting the
unitarity of these representations, one can write (\ref{fk.9}) in
the more familiar form
\begin{equation}\label{fk.10}
\int\d U\; U_\lambda{}^a{}_b\,U^*_\mu{}_d{}^c=
\frac{\vo}{\dim_\lambda}\,\delta_{\lambda\,\mu}\,\delta^a_d\,\delta^c_b.
\end{equation}
Using this orthogonality relation, one can obtain an orthogonality
relation between the characters of the group:
\begin{equation}\label{fk.11}
\int\d
U\;\chi_\lambda(U)\,\chi_\mu(U^{-1})=\vo\,\delta_{\lambda\,\mu},
\end{equation}
or
\begin{equation}\label{fk.12}
\int\d
U\;\chi_\lambda(U)\,\chi^*_\mu(U)=\vo\,\delta_{\lambda\,\mu},
\end{equation}
where
\begin{equation}\label{fk.13}
\chi_\lambda(U):=U_\lambda{}^a{}_a.
\end{equation}

The delta distribution is defined through
\begin{equation}\label{fk.14}
\int\d U\;\delta(U)\,f(U):=f(\id),
\end{equation}
where $\id$ is the identity element of the group; we notice that
as usual the delta picks up the value of the function at the
origin, $U=\id$. It is easy to see that this delta distribution is
invariant under similarity transformations, as well as inversion
of the argument:
\begin{align}\label{fk.15}
\delta(V\,U\,V^{-1})&=\delta(U),\cr \delta(U^{-1})&=\delta(U).
\end{align}
The first relation shows that if the argument of the delta is a
product of group elements, then any cyclic permutation of these
elements leaves the delta unchanged.

The regular representation of the group is defined through
\begin{equation}\label{fk.16}
U_\reg\,\e(V):=\e(U\,V),
\end{equation}
from which it is seen that the matrix element of this linear
operator is
\begin{equation}\label{fk.17}
U_\reg(W,V)=\delta(W^{-1}\,U\,V).
\end{equation}
This shows that the trace of the regular representation is
proportional to the delta distribution:
\begin{align}\label{fk.18}
\chi_\reg(U)=&\int\d V\;U_\reg(V,V),\cr =&\vo\,\delta(U).
\end{align}
So the delta distribution can be expanded in terms of the matrix
functions (in fact in terms of the characters of irreducible
representations). The result is
\begin{equation}\label{fk.19}
\delta(U)=\sum_\lambda\frac{\dim_\lambda}{\vo}\,\chi_\lambda(U),
\end{equation}
or
\begin{align}\label{fk.20}
\delta(U\,V^{-1})=&\sum_\lambda\frac{\dim_\lambda}{\vo}\,
U_\lambda{}^a{}_b\,V^{-1}_\lambda{}^b{}_a,\cr
=&\sum_\lambda\frac{\dim_\lambda}{\vo}\,U_\lambda{}^a{}_b\,V^*_\lambda{}_a{}^b.
\end{align}
This shows that the other functions are also expandable in terms
of the matrix functions:
\begin{equation}\label{fk.21}
f(U)=\sum_\lambda\frac{\dim_\lambda}{\vo}\,
U_\lambda{}^a{}_b\,f_\lambda{}_a{}^b,
\end{equation}
where
\begin{align}\label{fk.22}
f_\lambda{}_a{}^b:=&\int\d V\;V^{-1}_\lambda{}^b{}_a\,f(V),\cr
=&\int\d V\;V^*_\lambda{}_a{}^b\,f(V).
\end{align}
Using this and (\ref{fk.8}), one arrives at
\begin{equation}\label{fk.23}
(f\,g)_\lambda{}_a{}^b=f_\lambda{}_a{}^c\,g_\lambda{}_c{}^b.
\end{equation}

Next, one can define an inner product on the group algebra.
Defining
\begin{equation}\label{fk.24}
\langle\e(U),\e(V)\rangle:=\delta(U^{-1}\,V),
\end{equation}
and demanding that the inner product be linear with respect to its
second argument and antilinear with respect to its first argument,
one arrives at
\begin{align}\label{fk.25}
\langle f,g\rangle=&\int\d U\;f^*(U)\,g(U),\cr
=&\sum_\lambda\frac{\dim_\lambda}{\vo}\,f^*_\lambda{}^a{}_b\,g_\lambda{}_a{}^b.
\end{align}

Finally, one defines a star operation through
\begin{equation}\label{fk.26}
f^\c(U):=f^*(U^{-1}).
\end{equation}
This is in fact equivalent to definition of the star operation in
the group algebra by
\begin{equation}\label{fk.27}
[\e(U)]^\c:=\e(U^{-1}).
\end{equation}
It is then easy to see that
\begin{align}\label{fk.28}
(f\,g)^\c=&g^\c\,f^\c,\\ \label{fk.29} \langle f, g\rangle=&
(f^\c\,g)(\id).
\end{align}
Here a note is in order. While the results of this section were
obtained for compact groups, in some cases  compactness is not
necessary. It is easy to see that, provided (\ref{fk.4}) holds,
(\ref{fk.6})-(\ref{fk.8}), (\ref{fk.14})-(\ref{fk.17}),
(\ref{fk.24}), the first equality in (\ref{fk.25}), and
(\ref{fk.26})-(\ref{fk.29}) are still true, even if the group is
noncompact.

\subsection{Field theory}
Based on the calculational tools presented in the previous
subsection, here we can present the construction of a field theory
on a noncommutative space, the commutation relations of which are
those of a compact Lie group. In this work we consider the
simplest case: the scalar theory. To avoid explicit calculus on
such a noncommutative space, everything is defined on the momentum
space. This space is commutative and one can attribute
well-defined (local) coordinates to it, so that ordinary
differential and integral calculus (on manifolds) can be performed
on it. As far as observables of field theories are concerned, this
momentum representation is sufficient.

To give motivation for the particular form of the action that is
going to be written for a real scalar field, we first consider the
real scalar field on an ordinary $\mathbb{R}^D$ space. To be
consistent with the notation used throughout this paper, the
Fourier transform (only on space) of the field is denoted by
$\phi$, while the field itself is denoted by $\tilde\phi$. So we
have
\begin{equation}\label{fk.30}
\tilde\phi(\mathbf{r})=\int\frac{\d^D k}{(2\,\pi)^D}
\;\phi(\mathbf{k})\,\exp(\ir\,\mathbf{r}\cdot\mathbf{k}).
\end{equation}
An action for a scalar field is
\begin{equation}\label{fk.31}
S=\int\d t\,\d^D
r\;\left\{\frac{1}{2}\,\left[\dot{\tilde\phi}(\mathbf{r})\,
\dot{\tilde\phi}(\mathbf{r})+ \tilde\phi(\mathbf{r})\,\tilde
O(\nabla)\,\tilde\phi(\mathbf{r})\right]-\sum_{j=3}^n\frac{g_j}{j!}\,
[\tilde\phi(\mathbf{r})]^j\right\},
\end{equation}
where the $g_j$ are constants and $\tilde O(\nabla)$ is a
differential operator. This action is translation invariant, that
is, invariant under the transformations
\begin{equation}\label{fk.32}
\tilde\phi(\mathbf{r})\to{\tilde\phi}'(\mathbf{r}):=
{\tilde\phi}(\mathbf{r}-\mathbf{a}),
\end{equation}
where $\mathbf{a}$ is constant.

One can write the action (\ref{fk.31}) and the transformation
(\ref{fk.32}) in terms of the Fourier transforms:
\begin{align}\label{fk.33}
S=\int\d t\;\Bigg\{\frac{1}{2}&\int\frac{\d^D k_1\,\d^D
k_2}{(2\,\pi)^{2\,D}}\;[\dot\phi(\mathbf{k}_1)\,
\dot\phi(\mathbf{k}_2)\cr +&
\phi(\mathbf{k}_1)\,O(\mathbf{k}_2)\,\phi(\mathbf{k}_2)]\,
[(2\,\pi)^D\,\delta(\mathbf{k}_1+\mathbf{k}_2)]\cr
-&\sum_{j=3}^n\frac{g_j}{j!}\int\left[\prod_{l=1}^j\frac{\d^D
k_l\;\phi(\mathbf{k}_l)}{(2\,\pi)^D}\right]\;
[(2\,\pi)^D\,\delta(\mathbf{k}_1+\cdots+\mathbf{k}_j)]\Bigg\},
\end{align}
and
\begin{equation}\label{fk.34}
\phi(\mathbf{k})\to\phi'(\mathbf{k}):=
\exp(-\ir\,\mathbf{k}\cdot\mathbf{a})\,\phi(\mathbf{k}).
\end{equation}

Considering the space of the $\mathbf{k}$ as a group
($\mathbb{R}^D$), one notices that $(\d^D k)/(2\,\pi)^D$ is the
measure of this group that is invariant under right translation,
left translation, and inversion. It is not normalizable in the
sense of (\ref{fk.5}), as this group is not compact. One also
notices that $\exp(-\ir\,\mathbf{k}\cdot\mathbf{a})$ is nothing
but the representation $\mathbf{a}$ of the group element
corresponding to the coordinates $\mathbf{k}$. As this
representation is one dimensional,
$\exp(-\ir\,\mathbf{k}\cdot\mathbf{a})$ is also the determinant of
this representation.

Now we come to the case on fuzzy space.
A real scalar field $\phi$ is defined as a real member of the group algebra:
\begin{equation}\label{fk.35}
\phi^\c=\phi.
\end{equation}
In analogy with the action on ordinary space, one may suggest the
action
\begin{align}\label{fk.36}
S=\int\d t\Bigg\{\frac{1}{2}&\int\d U_1\,\d
U_2\;\left[\dot\phi(U_1)\,\dot\phi(U_2)+\int\d
U\;\phi(U_1)\,O(U_2,U)\,\phi(U)\right]\,\delta(U_1\,U_2)\cr
-&\sum_{j=3}^n\frac{g_j}{j!}\int\left[\prod_{l=1}^j\d
U_l\;\phi(U_l)\right]\,\delta(U_1\cdots U_j)\Bigg\}.
\end{align}
where $g_j$ are constants and $O$ is a linear operator from the
group algebra to the group algebra. For the action on the ordinary
space, one has
\begin{equation}\label{fk.37}
O(\mathbf{k}_2,\mathbf{k})\propto\delta(\mathbf{k}_2-\mathbf{k}).
\end{equation}
In analogy with that, we take
\begin{equation}\label{fk.38}
O(U_2,U)=O(U)\,\delta(U_2\,U^{-1}).
\end{equation}
From now on, it is assumed that this is the case. So
\begin{align}\label{fk.39}
S=\int\d t\Bigg\{\frac{1}{2}&\int\d U_1\,\d
U_2\;\left[\dot\phi(U_1)\,\dot\phi(U_2)+
\phi(U_1)\,O(U_2)\,\phi(U_2)\right]\,\delta(U_1\,U_2)\cr
-&\sum_{j=3}^n\frac{g_j}{j!}\int\left[\prod_{l=1}^j\d
U_l\;\phi(U_l)\right]\,\delta(U_1\cdots U_j)\Bigg\},
\end{align}
A simple choice for $O$ is
\begin{equation}\label{fk.40}
O(U)=c\,\chi_\lambda(U+U^{-1}-2\,\id)-m^2,
\end{equation}
where $\lambda$ is a representation of the group, and $c$ and $m$
are constants. An argument for the plausibility of this choice is
the following. Consider a Lie group and a group element near its
identity, so that
\begin{align}\label{fk.41}
U_\lambda=&~\exp(\tilde k^a\,T_{a\,\lambda}),\cr
\approx&~\id_\lambda+\tilde k^a\,T_{a\,\lambda}+\frac{1}{2}\,
(\tilde k^a\,T_{a\,\lambda})^2,
\end{align}
where $T_a$ are the generators of the group. One has
\begin{equation}\label{fk.42}
O(U)\approx c\,\chi_\lambda(T_a\,T_b)\,\tilde k^a\,\tilde k^b-m^2,
\end{equation}
which is a constant plus a bilinear form in $\tilde{\mathbf{k}}$,
just as was expected for an ordinary scalar field. In fact, if one
introduces a small constant $\ell$ so that $\tilde k$ is
proportional to $\ell$, and $c$ is proportional to $\ell^{-2}$,
then in the limit $\ell\to 0$ the expression (\ref{fk.42}) is
exactly equal to a constant plus a bilinear form.

An action of the form (\ref{fk.39}) with the choice (\ref{fk.40})
 also has a symmetry under
\begin{equation}\label{fk.43}
\phi(U)\to\phi(V\,U\,V^{-1}),
\end{equation}
where $V$ is an arbitrary member of the group.

One can write the action (\ref{fk.39}) in terms of the Fourier
transform of the field in time:
\begin{equation}\label{fk.44}
\phi(t,U)=:\int\frac{\d\omega}{2\,\pi}\;\exp(-\ir\,\omega\,t)
\,\check\phi(\omega,U),
\end{equation}
to arrive at
\begin{align}\label{fk.45}
S=&\frac{1}{2}\int\frac{\d\omega_1\,\d
U_1}{2\,\pi}\,\frac{\d\omega_2\,\d U_2}{2\,\pi}
\;\left[-\omega_1\,\omega_2\,\check\phi(U_1)\,\check\phi(U_2)+
\check\phi(U_1)\,O(U_2)\,\check\phi(U_2)\right]\cr
&\times[2\,\pi\,\delta(\omega_1+\omega_2)\,\delta(U_1\,U_2)]\cr
-&\sum_{j=3}^n\frac{g_j}{j!}\int\left[\prod_{l=1}^j
\frac{\d\omega_l\,\d U_l}{2\,\pi}\;\check\phi(U_l)\right]
\,[2\,\pi\,\delta(\omega_1+\cdots+\omega_j)\,\delta(U_1\cdots
U_j)].
\end{align}
The first two terms represent a free action, with the propagator
\begin{equation}\label{fk.46}
\check\Delta(\omega,U):=\frac{\ir\,\hbar}{\omega^2+O(U)}.
\end{equation}
Putting the denominator of this propagator equal to zero gives the
relation between $\omega$ and $U$ for free particles (the
mass-shell condition). The third term contains interactions. Any
Feynman graph would consist of propagators and $j$-line vertices
to which one assigns
\begin{equation}\label{fk.47}
V_j:=\frac{g_j}{\ir\,\hbar\,j!}\,2\,\pi\,
\delta(\omega_1+\cdots+\omega_j)\,\sum_{\Pi}\delta(U_{\Pi(1)}\cdots
U_{\Pi(j)}),
\end{equation}
where the summation runs over all $j$-permutations. In practice,
as we will see later, due to cyclic symmetry of arguments of the
$\delta$ functions mentioned earlier, permutations that are
different up to a cyclic change just come in the sum with a proper
weight. Also, for any internal line there is an integration over
$U$ and $\omega$, with the measure $\d\omega\,\d U/(2\,\pi)$. As
the group is assumed to be compact, the integration over the group
is integration over a compact volume. Hence there would be no
UV-divergences.

It is worth to mention a crucial difference between the way that
$\delta$ functions appear in our model and in models defined on
ordinary spaces. Here, as mentioned above, each possible ordering
of legs of a vertex comes with a different $\delta$, except the
cases that two orderings are different up to a cyclic permutation.
This is in contrast to models on ordinary space, in which all
possible orderings have the common factor of one single
$\delta\big( \sum \mathbf{k}_i\big)$, representing the momentum
conservation in that vertex.

Similar to the above observation, $\delta$ functions appear in
theories defined on $\kappa$-deformed spaces, as pointed out in the
Introduction . In these theories, the ordinary summation of
momenta in each vertex is replaced with a new summation rule,
occasionally called a dotted sum ($\dot{+}$) \cite{amelino}. This
new sum, contrary  to an  ordinary sum, is non-Abelian, and as a
consequence, the $\delta$ function coming with each possible
ordering of the legs are different \cite{amelino,kappa}.

One can compare this model to a field theory on a group manifold.
In the latter model, the integration in (\ref{fk.36}) or
(\ref{fk.39}) would be over the position, not over the momenta,
and the operator $O$ would be differentiation with respect to the
coordinates. In a model on a group manifold, the position
coordinates are still commuting but the momenta are not. Here the
situation is reversed, and this is not only a matter of
convenience. The operator $O$ determines which model is being
investigated: it is algebraic in terms of the momenta and
differentiation in terms of the position. For models on group
manifolds with compact groups, there would be no infrared (IR)
divergences while here there is no UV-divergence. The fact that
for a noncommutative geometry based on  the Lie groups the momenta
are still commuting is the reason that here the momentum picture
has been preferred to the position picture.

\subsection{An example: the group SU(2)}
For the group SU(2), one has
\begin{equation}\label{fk.48}
f^a{}_{b\,c}=\epsilon^a{}_{b\,c}.
\end{equation}
A group element $U$ can be characterized by the coordinates
$(k^1,k^2,k^3)$ such that
\begin{equation}\label{fk.49}
U=\exp(\ell\,k^a\,T_a),
\end{equation}
where $\ell$ is a constant. The invariant measure is
\begin{equation}\label{fk.50}
\d U=\frac{\sin^2(\ell\,k/2)}{(\ell\, k/2)^2}\,\frac{\d^3
k}{(2\,\pi)^3},
\end{equation}
where
\begin{equation}\label{fk.51}
k:=\left(\delta_{a\,b}\,k^a\,k^b\right)^{1/2}.
\end{equation}
The reason for this particular choice of normalization is that for
small values of $k$, (\ref{fk.50}) reduces to the integration
measure corresponding to the ordinary space. The integration
region for the coordinates is
\begin{equation}\label{fk.52}
k\leq\frac{2\,\pi}{\ell}.
\end{equation}

In the small-$k$ limit, one also has
\begin{equation}\label{fk.53}
\delta(U_1\,\cdots\,U_l)\approx (2\,\pi)^3\, \delta^3
(\mathbf{k}_1+\cdots+\mathbf{k}_l),
\end{equation}
which ensures an approximate momentum conservation. The exact
conservation law, however, is that at each vertex the product of
incoming group elements should be unity. For the case of a
three-leg vertex, one can write this condition as
\begin{equation}\label{fk.54}
\exp(\ell\,k_1^a\,T_a)\,\exp(\ell\,k_2^a\,T_a)\,
\exp(\ell\,k_3^a\,T_a)=1,
\end{equation}
or a similar condition in which $\mathbf{k}_1$ is replaced by
$\mathbf{k}_2$ and vice versa. One has
\begin{equation}\label{fk.55}
\exp(\ell\,k_1^a\,T_a)\,\exp(\ell\,k_2^a\,T_a)=:
\exp[\ell\,\gamma^a(\mathbf{k}_1,\mathbf{k}_2)\,T_a],
\end{equation}
where the function $\boldsymbol{\gamma}$ enjoys the properties
\begin{align}\label{fk.56}
\boldsymbol{\gamma}[\mathbf{k}_1,
\boldsymbol{\gamma}(\mathbf{k}_2,\mathbf{k}_3)]=&
\boldsymbol{\gamma}[\boldsymbol{\gamma}(\mathbf{k}_1,\mathbf{k}_2),
\mathbf{k}_3],\\ \label{fk.57}
\boldsymbol{\gamma}(-\mathbf{k}_1,-\mathbf{k}_2)=&
-\boldsymbol{\gamma}(\mathbf{k}_2,\mathbf{k}_1),\\ \label{fk.58}
\boldsymbol{\gamma}(\mathbf{k},-\mathbf{k})=&0.
\end{align}
Therefore, (\ref{fk.54}) becomes one of the three equivalent forms
\begin{align}\label{fk.59}
\mathbf{k}_3=&-\boldsymbol{\gamma}(\mathbf{k}_1,\mathbf{k}_2),\cr
\mathbf{k}_2=&-\boldsymbol{\gamma}(\mathbf{k}_3,\mathbf{k}_1),\cr
\mathbf{k}_1=&-\boldsymbol{\gamma}(\mathbf{k}_2,\mathbf{k}_3).
\end{align}
The explicit form of $\boldsymbol{\gamma}$ is obtained from
\begin{align}\label{fk.60}
\cos\frac{\ell\,\gamma}{2}=&
~\cos\frac{\ell\,k_1}{2}\,\cos\frac{\ell\,k_2}{2}-
\frac{\mathbf{k}_1\cdot\mathbf{k}_2}{k_1\,k_2}\,
\sin\frac{\ell\,k_1}{2}\,\sin\frac{\ell\,k_2}{2},\cr
\frac{\gamma^a}{\gamma}\,\sin\frac{\ell\,\gamma}{2}=&
~\epsilon^a{}_{b\,c}\,\frac{k_1^b\,k_2^c}{k_1\,k_2}\,
\sin\frac{\ell\,k_1}{2}\,\sin\frac{\ell\,k_2}{2}\cr &+
\frac{k_1^a}{k_1}\,\sin\frac{\ell\,k_1}{2}\,\cos\frac{\ell\,k_2}{2}+
\frac{k_2^a}{k_2}\,\sin\frac{\ell\,k_2}{2}\,\cos\frac{\ell\,k_1}{2}.
\end{align}
It is easy to see that in the limit $\ell\to 0$,
$\boldsymbol{\gamma}$ tends to $\mathbf{k}_1+\mathbf{k}_2$, as
expected.

The choice (\ref{fk.40}) for $O$ turns to be
\begin{equation}\label{fk.61}
O=2\,c\,
\left\{\frac{\displaystyle{\sin\left[\left(s+\frac{1}{2}\right)\,\ell\,k\right]}}
{\displaystyle{\sin\frac{\ell\,k}{2}}}-(2\,s+1)\right\}-m^2,
\end{equation}
where $s$ is the spin of the representation. For small values of
$k$, this is turned to
\begin{equation}\label{fk.62}
O\approx-c\,\frac{s\,(s+1)\,(2\,s+1)}{3}\,(\ell k)^2-m^2,\qquad
(\ell\,k)\ll 1.
\end{equation}
One chooses $c$ so that in the small-$k$ limit $O$ takes the
ordinary form of the propagator inverse:
\begin{equation}\label{fk.63}
O\approx-k^2-m^2,\qquad (\ell\,k)\ll 1.
\end{equation}
Choosing
\begin{equation}\label{fk.64}
c=\frac{3}{s\,(s+1)\,(2\,s+1)\,\ell^2},
\end{equation}
the propagator becomes
\begin{equation}\label{fk.65}
\check\Delta(\omega,\mathbf{k})=\frac{\ir\,\hbar}
{\omega^2+\displaystyle{\frac{6}{s\,(s+1)\,(2\,s+1)\,\ell^2}}\,
\left\{\frac{\displaystyle{\sin\left[\left(s+\frac{1}{2}\right)\,
\ell\,k\right]}}
{\displaystyle{\sin\frac{\ell\,k}{2}}}-(2\,s+1)\right\}-m^2}.
\end{equation}
It is easy to see that in the limit $\ell\to 0$, the usual
commutative propagator is recovered.

Similar things hold for the group SO(3). One only has to replace
the integration region by
$$
k\leq\frac{\pi}{\ell}.\eqno{(52')}
$$
A consequence of the compactness of the momentum space is that
field theories based of spaces with Lie group fuzziness
corresponding to compact groups are free from UV-divergences. The
above restriction on the integration region in momentum space, as
well as the UV-finiteness of theory, are very similar to those one
has in theories defined on lattices. This would be no surprise for
this behavior once one mentions that the eigenvalues of the space
coordinates are discrete as a consequence of the coordinates
satisfying the SU(2) or SO(3) algebras, and in general that of a
compact Lie group. There are, however, differences between such
theories and theories based on space lattices: In the latter
theories there are no continuous space symmetries, while in the
former one there are (rotation in the case of SO(3) or SU(2)); in
the former case it is not possible to determine all position
operators simultaneously, while in the latter case it is; and in
the latter case the positions are discrete, while in the former
case the position eigenvalues are discrete.

The UV-finiteness of the model is reminiscent of the old
expectation that in noncommutative spaces the theory might be free
from the divergences caused by the short distance behavior of
physical quantities. In this sense noncommutative theories based
on compact groups resemble ordinary (commutative theories) with a
momentum cutoff. It would be interesting to mention the fate of
the UV/IR mixing phenomena \cite{MRS}. As a generic property of
models defined on canonical noncommutative spaces,
see (\ref{fk.1}), certain combinations of external momenta and the
noncommutativity parameter $\theta$ may appear as a dynamical
cutoff in momentum space. For example, in two-external leg
diagrams of $\phi^4$ theory, the combination $(p\circ p)^{-1/2}$
with $p\circ p:= (p^\mu\theta_{\mu\nu}^2p^\nu)$ acts as a cutoff,
causing the contribution of the so called non-planar diagram to be
UV-finite \cite{MRS}. In the extreme IR limit of external momenta
($p\to 0$), this cutoff tends to infinity and the result diverges.
In such a case, in the IR limit of the theory the UV-divergences
of the commutative (ordinary) theory are restored. This is the so
called UV/IR mixing. If the noncommutative theory had been based
on a commutative theory with a momentum cutoff, there would be no
UV-divergence and no UV/IR mixing.

Theories discussed here are free from UV-divergences, as the
momentum space is compact. In this sense, they are based on
commutative theories with a momentum cutoff. Hence there is no
UV-divergence in the original theory to be restored in some IR limit,
and there is no room for UV/IR mixing.

\section{Amplitudes}
In this section the basic elements for calculation of a transition
amplitude, including the construction of initial and final states,
the proper normalization of the states, and the relevant
kinematical factors are presented.

\subsection{Fock space and initial/final states}
According to the  previous section, the free sector of the
Lagrangian in the momentum space is given by
\begin{equation}\label{fk.66}
L_{\rm free}=\frac{1}{2}\int\d
U\,\;\left[\dot\phi(U^{-1},t)\,\dot\phi(U,t) +
\phi(U^{-1},t)\,O(U)\,\phi(U,t)\right],
\end{equation}
from which one obtains the canonical field momenta
\begin{equation}\label{fk.67}
\Pi(U,t)= \dot\phi(U^{-1},t).
\end{equation}
The equal-time canonical commutation relations are
\begin{align}\label{fk.68}
\left[\phi(U,t),\Pi(V,t)\right]&=\ir \hbar\, \delta(U\,V^{-1}),\cr
\left[\phi(U,t),\phi(V,t)\right]&=0,\cr
\left[\Pi(U,t),\Pi(V,t)\right]&=0.
\end{align}
As usual one might express the dynamical variables in terms of
positive and negative frequency components:
\begin{equation}\label{fk.69}
\phi(U,t)=\sqrt{\frac{\hbar}{2\omega}}\,\left[a(U)\,\exp(-\ir\,\omega\,
t) + a^\dagger(U^{-1})\,\exp(\ir\,\omega\, t) \right],
\end{equation}
from which one finds
\begin{align}\label{fk.70}
\left[a(U),a^\dagger(V)\right]&=\delta(UV^{-1}),\cr
\left[a(U),a(V)\right]&=0,\cr
\left[a^\dagger(U),a^\dagger(V)\right]&=0.
\end{align}
One defines the vacuum-state through
\begin{align}\label{fk.71}
a(U)\,|0\rangle&=0,\qquad\forall~ U,\cr \langle 0|0\rangle&=1.
\end{align}
The multi-particle states with given momenta, being a basis of the
Fock space of theory, are constructed as
\begin{equation}\label{fk.72}
|(U_1,n_1); (U_2,n_2);
\cdots\rangle:=\frac{[a^\dagger(U_1)]^{n_1}}{\sqrt{n_1
!}}\,\frac{[a^\dagger(U_2)]^{n_2}}{\sqrt{n_2 !}}\cdots |0\rangle.
\end{equation}
Equetions(\ref{fk.71}) and (\ref{fk.72}) also give the
normalization of multi-particle states. For example,
\begin{align}\label{fk.73}
\langle U | V \rangle &= \delta(U^{-1} V),\cr \langle U | U
\rangle &= \delta(\id).
\end{align}
Of course, the right-hand side of the latter is infinite. But this
is similar to the case of ordinary space. In the case of ordinary
space, the left-hand side is finite if and only if the volume of
the system is finite. In that case the left-hand side is equal to
the volume of the system. One can keep the volume of the system
finite and do calculations up to the point where this volume is no
longer there in the observables, and then send the volume to
infinity. The same thing is possible here too. In this case,
instead of talking about the finiteness of the volume one takes a
finite number of representations of the group. Again one does the
calculations until this {\em volume} in the right-hand side
disappears, and then sends the upper limit on the representations
to infinity. The overall result is that one takes $\delta(\id)$ as
the volume of the system and deals with it like a finite number
(in the intermediate stages of the calculations). In the final
result, however, there should not be any $\delta(\id)$.

\subsection{{\itshape S}-matrix and transition amplitudes}
An element of $S$-matrix, which represents the transition from the
initial state $\mathrm{i}$ to the final state $\mathrm{f}$, would
come in the general form
\begin{equation}\label{fk.74}
S_{\mathrm{f\,i}}=\delta_{\mathrm{f\,i}}+ T_{\mathrm{f\,i}},
\end{equation}
where the matrix elements of $T$ come from the interaction terms.
In the case of commutative space, $T_{\mathrm{f\,i}}$ contains a
delta distribution corresponding to energy conservation and
another delta distribution corresponding to momentum conservation.
It also contains (corresponding to each incoming or outgoing
particle) a factor $\sqrt{\hbar/(2\,\omega)}$ (coming from the
expression of the field in terms of creation and annihilation
operators) as well as a normalization factor
$\sqrt{1/\mathcal{V}}$ (where $\mathcal{V}$ is the volume of the
space). One then has
\begin{align}\label{fk.75}
T_{\mathrm{f\,i}}=:&2\,\pi\,\delta\left(\sum_j\omega_{\mathrm{f}\,j}-
\sum_l\omega_{\mathrm{i}\,l}\right)\,
(2\,\pi)^D\,\delta\left(\sum_j\mathbf{k}_{\mathrm{f}\,j}-
\sum_l\mathbf{k}_{\mathrm{i}\,l}\right)\cr &\times\,\prod_j
\sqrt{\frac{\hbar}{2\,\omega_{\mathrm{f}\,j}\,\mathcal{V}}}\,
\prod_l\sqrt{\frac{\hbar}{2\,\omega_{\mathrm{i}\,l}\,\mathcal{V}}}\,
~\tilde M_{\mathrm{f\,i}},
\end{align}
for ordinary space.In the case of noncommutative space, instead of
$\mathcal{V}$, one has $\delta(\id)$, and instead of the delta
distribution corresponding to momentum conservation one has a
delta distribution of a product of group elements corresponding to
incoming and outgoing particles. Contrary to the case of ordinary
space, however, the order of these group elements in the delta
distribution is important. In this case one has
\begin{align}\label{fk.76}
T_{\mathrm{f\,i}}=:&2\,\pi\,\delta\left(\sum_j\omega_{\mathrm{f}\,j}-
\sum_l\omega_{\mathrm{i}\,l}\right)\cr &\times\,\prod_j
\sqrt{\frac{\hbar}{2\,\omega_{\mathrm{f}\,j}\,\delta(\id)}}\,
\prod_l\sqrt{\frac{\hbar}{2\,\omega_{\mathrm{i}\,l}\,\delta(\id)}}\,
~\mathcal{M}_{\mathrm{f\,i}},
\end{align}
where
\begin{equation}\label{fk.77}
\mathcal{M}_{\mathrm{f\,i}}=\sum_{\Pi}M^{\Pi}_{\mathrm{f\,i}}\,
\delta(U^{\Pi}).
\end{equation}
Here $U^{\Pi}$ is a symbolic notation meaning a product of group
elements corresponding to outgoing particles, the inverse of group
elements corresponding to incoming particles, and possibly group
elements corresponding the loops integrated. The order of these
elements is symbolically determined by $\Pi$.

$T_{\mathrm{f\,i}}$ is the amplitude of the transition. The
probability of transition is the square of its modulus times the
number of final states:
\begin{equation}\label{fk.78}
p_{\,\mathrm{i}\to\mathrm{f}}=|T_{\mathrm{f\,i}}|^2\,\prod_j
[\delta(\id)\,\d U_{{\mathrm f}\,j}].
\end{equation}
The factors $\delta(\id)$ in the number of final states cancel the
factors $\delta(\id)$ corresponding to outgoing particles in
$|T_{\mathrm{f\,i}}|^2$. There remains the factors $\delta(\id)$
corresponding to incoming particles. In $|T_{\mathrm{f\,i}}|^2$,
each term contains a product of two delta distribution of
appropriate group elements, $\delta(U^{\Pi})\,\delta(U^{\Pi'})$.
If $(U^{\Pi}=\id)$ is equivalent to $(U^{\Pi'}=\id)$, then one can
write $\delta(U^{\Pi})\,\delta(U^{\Pi'})$ as
$\delta(U^{\Pi})\,\delta(\id)$. This means that in
$|T_{\mathrm{f\,i}}|^2$ divided by $\delta(\id)$, only those terms
survive that come from $[\delta(U^{\Pi})]^2$. That is,
\begin{equation}\label{fk.79}
|\mathcal{M}_{\mathrm{f\,i}}|^2\to
\delta(\id)\,\sum_{\Pi}|M^{\Pi}_{\mathrm{f\,
i}}|^2\,\delta(U^{\Pi}).
\end{equation}
Note the difference with the case of ordinary space. In that case
one would have $|\sum_{\Pi}M_{\mathrm{f\, i}}|^2$ instead of
$\sum_{\Pi}|M_{\mathrm{f\, i}}|^2$.

The rest is similar to the case of ordinary space. For a decay
process, $\delta(\id)$ in the right-hand side of (\ref{fk.79})
cancels the remaining $\delta(\id)$ coming from the normalization
of the state of the incoming particle. For a two-particle
collision, one has
\begin{equation}\label{fk.80}
\sigma\propto
p_{\mathrm{i}\to\mathrm{f}}\,\frac{1}{v_{\mathrm{rel}}\,\delta(\id)},
\end{equation}
where $v_{\mathrm{rel}}$ is the speed of the colliding particle
relative to the target, and $1/[\delta(\id)]$ is the density of
the colliding particles (one particle in a volume $\mathcal{V})$.
The factor $\delta(\id)$ in the right-hand side of the above
expression cancels the remaining $\delta(\id)$ in
$|T_{\mathrm{f\,i}}|^2$, so that at the end there remains no
factor of $\delta(\id)$, as expected.

In $|T_{\mathrm{f\,i}}|^2$, there is also a term
$[2\,\pi\,\delta(\omega_{\mathrm{f}}-\omega_{\mathrm{i}})]^2$,
which can be written as
$\mathcal{T}\,[2\,\pi\,\delta(\omega_{\mathrm{f}}-\omega_{\mathrm{i}})]$,
where $\mathcal{T}$ is the interaction time, which should be sent
to infinity. The transition rate is the probability divided by
$\mathcal{T}$. Therefore, in the rate the factor $\mathcal{T}$ is
cancelled, just as in the case of ordinary space.

These results can be summarized as
\begin{equation}\label{fk.81}
\d\Gamma=\frac{\hbar}{2\,\omega_{\mathrm{i}}}\,
2\,\pi\,\delta(\omega_{\mathrm{f}}-\omega_{\mathrm{i}})\,\left[
\sum_{\Pi}|M^{\Pi}_{\mathrm{f\, i}}|^2\,\delta(U^{\Pi})\right]\,
\prod_j\,\left(\frac{\hbar}{2\,\omega_{\mathrm{f}\,j}}\,\d
U_{\mathrm{f}\,j}\right),
\end{equation}
for the decay rate $\Gamma$, and
\begin{equation}\label{fk.82}
\d\sigma=\frac{1}{v_{\mathrm{rel}}}\,
\prod_{l=1}^2\,\left(\frac{\hbar}{2\,\omega_{\mathrm{i}\,l}}\right)\,
2\,\pi\,\delta(\omega_{\mathrm{f}}-\omega_{\mathrm{i}})\,\left[
\sum_{\Pi}|M^{\Pi}_{\mathrm{f\, i}}|^2\,\delta(U^{\Pi})\right]\,
\prod_j\,\left(\frac{\hbar}{2\,\omega_{\mathrm{f}\,j}}\,\d
U_{\mathrm{f}\,j}\right),
\end{equation}
for the cross section $\sigma$ in a two-particle collision.

Finally, let us address the relative speed $v_{\mathrm{rel}}$. In
the case of ordinary space, one defines the relative speed through
\begin{equation}\label{fk.83}
v_{\mathrm{rel}}:=\sqrt{\delta^{a\,b}
\,\frac{\partial\omega(\mathbf{k})}{\partial
k^a}\,\frac{\partial\omega(\mathbf{k})}{\partial k^b}},
\end{equation}
where
\begin{equation}\label{fk.84}
\mathbf{k}=\mathbf{k}_1-\mathbf{k}_2,
\end{equation}
and $\mathbf{k}_1$ and $\mathbf{k}_2$ are the momenta of incoming
particles. This speed does not change under exchanging particles 1
and 2, or under a rotation of the incoming momenta. In fact, as
$\omega$ depends on only the length of $\mathbf{k}$, one has
\begin{equation}\label{fk.85}
v_{\mathrm{rel}}=\frac{\d\omega}{\d|\mathbf{k}|}.
\end{equation}

In the case of a noncommutative space, one works most conveniently
with group elements instead of momenta. Instead of
$\delta^{a\,b}$, one could use the matrix elements of an invariant
two-form of the algebra. One could choose the coordinates so that
these elements become $\delta^{a\,b}$. Instead of
$(\mathbf{k}_1-\mathbf{k}_2)$, one could use $(U_1\,U_2^{-1})$, or
$(U_2^{-1}\,U_1)$, or their inverses. Instead of differentiation
with respect to $k^a$, one could use the action of
$X^{\mathrm{L}}_a$ or $X^{\mathrm{R}}_a$, as the left and right
invariant vector fields, respectively, whose actions at the origin
(the unit element of the group) is equal to differentiation with
respect to $k^a$. So, one would have
\begin{equation}\label{fk.86}
v_{\mathrm{rel}}:=\sqrt{\delta^{a\,b} \,\{[L_{X_a}(\omega)](U)\}
\,\{[L_{X_b}(\omega)](U)\}},
\end{equation}
where $L_X(\omega)$ means the action (Lie derivative) of the
vector field $X$ on the function $\omega$. As there are four
choices for $U$ and two choices for $X$, it seems that one should
choose between eight possible definitions for the relative speed.
The function $\omega$, however, is a class function, that is
\begin{equation}\label{fk.87}
\omega(V\,U\,V^{-1})=\omega(U),
\end{equation}
as $\omega^2$ is in fact ``~$-O(U)$~". By this, together with the
fact that $X_a$ are left or right invariant, and that
$\delta^{a\,b}$ is an invariant two-form, one can show that all
these choices lead to the same value for $v_{\mathrm{rel}}$. Even
more, one can in fact substitute $X_a(\omega)$ with the partial
derivative of $\omega$ with respect to $k^a$. Then, as $\omega$ is
a function of $|\mathbf{k}|=\sqrt{\delta_{a\,b}\,k^a\,k^b}$, it is
seen that (\ref{fk.85}) holds for the case of noncommutative
spaces as well. In fact,
\begin{equation}\label{fk.88}
v_{\mathrm{rel}}=\frac{\d\sqrt{-O(U)}}{\d|\mathbf{k}|}.
\end{equation}

\subsection{Examples}
In this subsection explicit expressions for the perturbative
expansion of field theory amplitudes in a space with SU(2)
fuzziness are discussed.

For the propagator, let us choose the representation
$s=\frac{1}{2}$ in (\ref{fk.65}):
\begin{equation}\label{fk.89}
\check\Delta(\omega,\mathbf{k})=\frac{\ir\,\hbar}
{\omega^2-\displaystyle{\frac{16}{\ell^2}\,\sin^2\frac{\ell\,
k}{4}}-m^2}.
\end{equation}
The reason for this choice is that it is the only representation
for which, on the mass shell, energy is an increasing function of
momentum. By this choice, one has for the relative velocity
\begin{equation}\label{fk.90}
v_{\mathrm{rel}}=\frac{\frac{2}{\ell}\,\sin\frac{\ell\,k}{2}}
{\sqrt{\frac{16}{\ell^2}\,\sin^2\frac{\ell\,k}{4}+m^2}}.
\end{equation}
We consider two types of interactions, the $\phi^3$ and $\phi^4$
interactions, which correspond to nonzero $g_3$ and $g_4$ in
(\ref{fk.39}).

\subsubsection{The three-particle interaction}
The fundamental vertex with three incoming legs 1, 2, and 3 is
\begin{equation}\label{fk.91}
V_3^{[123]}=\frac{g_3}{2\,\ir\,\hbar}\,2\,\pi\,
\delta(\omega_1+\omega_2+\omega_3)\,\left[\delta(U_1\,U_2\,U_3)+
\delta(U_1\,U_3\,U_2)\right].
\end{equation}
Now consider the scattering process $1+2\to 3+4$. At the tree
level, this process occurs via three diagrams (the s-, t-, and
u-channels). Each of these channels correspond to four types group
element delta functions. Of the twelve group element delta
functions, however, there are only six different delta functions,
each appearing in two of the three channels. The overall result
corresponding to (\ref{fk.77}) is then
\begin{align}\label{fk.92}
\mathcal{M}_{\mathrm{f\,i}}=\left(\frac{g_3}{2\,\ir\,\hbar}\right)^2\,
\{&[\check\Delta(\omega_{\mathrm{s}},\mathbf{k}_{\mathrm{s}})+
\check\Delta(\omega_{\mathrm{t}},\mathbf{k}_{\mathrm{t}})]\,
\delta(U_1\,U_2\,U_4^{-1}\,U_3^{-1})\cr
+&[\check\Delta(\omega_{\mathrm{s}},\mathbf{k}_{\mathrm{s}})+
\check\Delta(\omega_{\mathrm{t}},\mathbf{k}_{\mathrm{t}})]\,
\delta(U_1\,U_3^{-1}\,U_4^{-1}\,U_2)\cr
+&[\check\Delta(\omega_{\mathrm{s}},\mathbf{k}_{\mathrm{s}})+
\check\Delta(\omega_{\mathrm{u}},\mathbf{k}_{\mathrm{u}})]\,
\delta(U_1\,U_2\,U_3^{-1}\,U_4^{-1})\cr
+&[\check\Delta(\omega_{\mathrm{s}},\mathbf{k}_{\mathrm{s}})+
\check\Delta(\omega_{\mathrm{u}},\mathbf{k}_{\mathrm{u}})]\,
\delta(U_1\,U_4^{-1}\,U_3^{-1}\,U_2)\cr
+&[\check\Delta(\omega_{\mathrm{t}},\mathbf{k}_{\mathrm{t}})+
\check\Delta(\omega_{\mathrm{u}},\mathbf{k}_{\mathrm{u}})]\,
\delta(U_1\,U_3^{-1}\,U_2\,U_4^{-1})\cr
+&[\check\Delta(\omega_{\mathrm{t}},\mathbf{k}_{\mathrm{t}})+
\check\Delta(\omega_{\mathrm{u}},\mathbf{k}_{\mathrm{u}})]\,
\delta(U_1\,U_4^{-1}\,U_2\,U_3^{-1})\},
\end{align}
where,
\begin{align}\label{fk.93}
\omega_{\mathrm{s}}&:=\omega_1+\omega_2,\cr
\omega_{\mathrm{t}}&:=\omega_1-\omega_3,\cr
\omega_{\mathrm{u}}&:=\omega_1-\omega_4,
\end{align}
and
\begin{align}\label{fk.94}
U_{\mathrm{s}}&:=U_1\,U_2,\cr U_{\mathrm{t}}&:=U_1\,U_3^{-1},\cr
U_{\mathrm{u}}&:=U_1\,U_4^{-1}.
\end{align}
It is to be noted that sending $\ell$ to zero, while makes the
propagators equal to the commutative ones, does {\em not} make the
transition rate equal to the commutative one. The origin of this
difference, as pointed  out in the previous section, comes back to
the way of appearance of the $\delta$. Here, as pointed out
earlier, each possible ordering of legs of a vertex or diagram
comes with a different $\delta$, except the cases that two
orderings are the same up to a cyclic permutation. This is in
contrast to models on ordinary space, in which all possible
orderings have the common factor of one single $\delta\big( \sum
\mathbf{k}_i\big)$, representing the momentum conservation in that
vertex. So, in the present case, the set of available final states
is larger than the corresponding set in the commutative case. As
it is seen from the delta functions, for given $\mathbf{k}_1$,
$\mathbf{k}_2$, and $\mathbf{k}_3$, there is not only one, but
there are six possible values of $\mathbf{k}_4$. In the
commutative case, all these six values are the same, so that one
should add the amplitudes and then square the result. In the
present case, these are not the same, so that one should add the
squares, as one is calculating the transition probability to
different final states. The overall result in the present case,
apart from a multiplicative constant, is that the ratio of terms
containing a propagator squared to the terms containing the
product of two different propagator is one. The corresponding
ratio in the commutative case is one half. As mentioned in the
previous section, a similar observation has been made in theories
defined on $\kappa$-deformed spaces \cite{amelino,kappa}.

\subsubsection{The four-particle interaction}
The fundamental vertex with four incoming legs 1, 2, 3, and 4 is
\begin{align}\label{fk.95}
V_4^{[1234]}=\frac{g_4}{6\,\ir\,\hbar}\,&2\,\pi\,
\delta(\omega_1+\omega_2+\omega_3+\omega_4)\cr \times&\left[
\delta(U_1\,U_2\,U_3\,U_4) + \delta(U_1\,U_2\,U_4\,U_3)+
\delta(U_1\,U_3\,U_2\,U_4)\right.\cr
+&\left.\;\delta(U_1\,U_3\,U_4\,U_2)+ \delta(U_1\,U_4\,U_2\,U_3)+
\delta(U_1\,U_4\,U_3\,U_2)\right].
\end{align}
For the scattering process $1+2\to 3+4$, at the tree level there
is a single diagram. The overall result corresponding to
(\ref{fk.77}) is then
\begin{align}\label{fk.96}
\!\!\!\!\mathcal{M}_{\mathrm{f\,i}}=\frac{g_4}{6\,\ir\,\hbar}\,&\left[
\delta(U_1\,U_2\,U_3^{-1}\,U_4^{-1}) +
\delta(U_1\,U_2\,U_4^{-1}\,U_3^{-1})+
\delta(U_1\,U_3^{-1}\,U_2\,U_4^{-1})\right.\cr
+&\left.\;\delta(U_1\,U_3^{-1}\,U_4^{-1}\,U_2)+
\delta(U_1\,U_4^{-1}\,U_2\,U_3^{-1})+
\delta(U_1\,U_4^{-1}\,U_3^{-1}\,U_2)\right].
\end{align}
In above one may observe how the different ordering of legs in a
vertex come with different $\delta$, again just as the same
phenomena in the $\kappa$-deformed theories \cite{amelino,kappa}.

\section{Conclusion}
The structure of field theory transition amplitudes in a
three-dimensional space whose spatial coordinates are noncommutative and
satisfy the SU(2) Lie algebra were examined. In particular, the
basic notions for constructing the observables of the theory were
introduced. These include multi-particle states of the theory as a
basis of Fock space, an instruction for the proper normalization of
the kinematical factors associated with initial and final states
of observables, as well as the way one can introduce the relative
velocity between the initial states, appearing in the incident
flux of an observable. Subtleties related to the proper treatment
of the $\delta$-distributions in a $S$-matrix expansion of the
theory were discussed. Explicit examples were given for the
amplitudes of an interacting scalar field theory in the lowest
order of perturbation theory.
\\
\\
\textbf{Acknowledgement}:  This work was partially supported by
the research council of the Alzahra University.

\end{document}